\documentclass[aps,prb,twocolumn,groupedaddress,nofootinbib]{revtex4-1}
\usepackage{amsmath,bm}
\usepackage{graphicx}
\usepackage[bottom]{footmisc}

\newcommand{\tbf}[1]{{\textbf{#1}}}

\def\re{{\text{Re}\,}}
\def\im{{\text{Im}\,}}
\def\Tr{{\text{Tr}\,}}

\newcommand{\oln}[1]{{\overline{#1}}}

\begin{document}



\title{Specific heat and pairing of Dirac composite fermions in the half-filled Landau level}


\author{Nicholas Rombes and Sudip Chakravarty}
\affiliation{Mani L Bhaumik Institute for Theoretical Physics\\Department of Physics and Astronomy, University of California Los Angeles, Los Angeles, California 90095, USA}

\date{\today}

\begin{abstract}
A recent proposal argues that an alternate description of the half-filled Landau level is a theory of massless Dirac fermions. We examine the possibility of pairing of these Dirac fermions by numerically solving the {\em coupled} Eliashberg equations unlike our previous calculation [Wang and Chakravarty, Phys. Rev. B {\bf 94}, 165138 (2016)]. In addition, vertex corrections are calculated to be zero from the Ward identity. We find that pairing is possible in non-zero angular momentum channels; the only differences are minor numerical shifts. As before, the pairing leads to the gapped Pfaffian and anti-Pfaffian states. However, in our approximation scheme,  pairing is not possible in the putative €œ
particle-hole symmetric€ state for $\ell=0$ angular momentum. The specific heat at low temperatures of a system of massless Dirac fermions interacting with a transverse gauge field, expected to be relevant for the half-filled Landau level, is calculated. Using the Luttinger formula, it is found be $\propto T\ln T$ in the leading low temperature limit, due to the exchange of transverse gauge bosons. The result agrees with the corresponding one in the  nonrelativistic composite fermion theory of Halperin, Lee and Read of the half-filled Landau level.
\end{abstract}


\maketitle

\section{Introduction\label{intro}}
The nature of the half-filled Landau level has been a topic of interest for some time. Experiments indicate a peculiar metallic state when the lowest Landau level is half-filled, with a dip in the diagonal resistivity but no plateau in the Hall resistivity \cite{Jiang:1989de}, anomalous acoustic wave propagation \cite{Willett:1990fn}, and enhancement of the effective mass \cite{Du:1994wtw}. A description was given by Halperin, Lee, and Read (HLR), in which the correct degrees of freedom are ``composite fermions'' (CFs) interacting with an emergent ``statistical'' gauge field with a Chern-Simons term, which serves to attach two flux quanta to the original electron \cite{Halperin:1993tt}. In this theory, the CFs move in a reduced magnetic field, which vanishes at the mean-field level at half-filling. 

While the half-filled lowest Landau level, in the limit of practically infinite Landau level separation, can be equally well described as either electrons populating an empty Landau level or holes populating a full Landau level (particle-hole (PH) symmetry), there is no obvious way to make this symmetry apparent within the HLR theory. Thus, recently, a radical description of the half-filled Landau level was proposed, in which the CFs are now massless Dirac particles, and there is no Chern-Simons term for the emergent gauge field \cite{Son:2015gz}. In this theory, PH symmetry is explicitly incorporated at half filling. It is a matter of debate whether these two descriptions, the HLR description and the Dirac CF description, represent equivalent formulations of the half-filled Landau level \cite{Levin:2017fk,Wang:2017kx}.

In this work we accomplish two goals: firstly, we construct a pairing mechanism for the Dirac CFs, and show that pairing is possible (with minor differences from our previous work~\cite{Wang:2016bh}) in angular momentum channels apart from $\ell=0$, for which we do not find pairing to be possible. Secondly, we compute the low-temperature specific heat of the Dirac CFs, which {\em does not} differ from the corresponding result in HLR theory. The present Eliashberg  calculation involves solving both the coupled equations involving  the order parameter  and the Eliashberg-$Z$ factor. We furthermore make use of the Luttinger formula for the free energy~\cite{Luttinger:1960zz}. It was shown in Ref.~ \onlinecite{Curnoe:1998fo} that this expansion fails in general for interacting fermionic systems in 2D; however, it is valid here in at least the leading order because the vertex correction vanishes, as shown   from the Ward identity in Appendix~{C}. Thus the present approximation is on much firmer footing than in Ref.~\onlinecite{Wang:2016bh}.

\section{Model\label{model}}
The low-energy effective action for the Dirac CF is given by \cite{Son:2015gz}
\begin{equation}{\label{effectiveAction}}
  S_{\text{CF}}=\int d^3x\{i\oln\psi\gamma_\mu(\partial_\mu+ia_\mu)\psi+\frac{1}{4\pi}\epsilon^{\mu\nu\lambda}A_\mu\partial_\nu a_\lambda\},
\end{equation}
where $\{\gamma_0,\gamma_1,\gamma_2\}=\{\sigma_3,\sigma_1,\sigma_2\}$ are the Pauli matrices, $\oln\psi=\psi^\dagger\gamma_0$, and we have set $\hbar=v_F=1$. In this work, Greek indices run from 0 to 2 and Roman indices run from 1 to 2. This action describes massless, electrically neutral Dirac fermions that are charged under an emergent gauge field $a_\mu$. Differenting this action with respect to $a_0$, we see that 
\begin{equation}
  \oln\psi\gamma_0\psi=\frac{\nabla\times\tbf{A}}{4\pi}.
\end{equation}
The density of Dirac CFs is set by the physical external magnetic field, and is not the same as the density of physical electrons, in contrast to the HLR description. Differentiating with respect to $A_0$, we find
\begin{equation}{\label{electronDensity}}
  \rho_e'=\frac{\nabla\times\tbf{a}}{4\pi}.
\end{equation}
Since the emergent gauge field strength $b\equiv\nabla\times\tbf{a}$ should be zero at half-filling, we interpret $\rho_e'$ as the difference between the physical electron density and its value at half-filling: $\rho_e'=\rho_e-\rho_{\nu=1/2}$. Thus the strength of the emergent gauge field is set by the physical electron density.

It is shown in Ref.~\onlinecite{Wang:2016bh}, reproduced here in Appendix A for the sake of completeness, how to obtain an effective BCS interaction between Dirac CFs. The resulting potential is given by
\begin{equation}
  V_{\ell'}(i\Omega)=\alpha\int_{-\pi}^\pi\frac{d\theta}{2\pi}\frac{e^{i(\ell'-1)\theta}}{|\sin\frac{\theta}{2}|}\frac{2}{1+\alpha\frac{|\Omega|}{\sin^2\frac{\theta}{2}}}.
\end{equation}
Here $\ell'$ is the angular momentum channel of the \emph{scalar} order parameter (see Appendix A), $\Omega$ is the Matsubara frequency transfer between the interacting fermions, $\alpha=\frac{\epsilon_rv_F}{e^2}$ is the effective coupling constant of the Dirac CFs, and both $V_{\ell'}$ and $\Omega$ are measured in units of the Fermi energy; the background dielectric constant is $\epsilon_{r}$. This potential serves as the kernel for the zero-temperature, imaginary axis Eliashberg equations:
\begin{equation}{\label{eliashberg1}}
\begin{split}
  \phi_{\ell'}(i\omega)=-\int_{-\infty}^\infty\frac{d\nu}{2\pi}&V_{\ell'}(i\omega-i\nu) \\
  &\times\frac{\phi_{\ell'}(i\nu)}{\sqrt{(\nu Z(i\nu))^2+|\phi_{\ell'}(i\nu)|^2}}
\end{split}
\end{equation}
\begin{equation}{\label{eliashberg2}}
\begin{split}
  [1-Z(i\omega)]\omega=\int_{-\infty}^\infty\frac{d\nu}{2\pi}&V_{\ell'=1}(i\omega-i\nu) \\
  &\times\frac{\nu Z(i\nu)}{\sqrt{(\nu Z(i\nu))^2+|\phi_{\ell'}(i\nu)|^2}}.
\end{split}
\end{equation}
Here $Z(i\omega)$ is the mass renormalization factor, $\Delta_{\ell'}(i\omega)\equiv\phi_{\ell'}(i\omega)/Z(i\omega)$ is the gap function, and we interpret $\Delta_{\ell'}(0)$ as the physical gap at $T=0$ where the frequencies form a continuum. Our goal will be to numerically solve these coupled integral equations.

\section{Results\label{results}}
\subsection{Eliashberg Equations}
Here we present our numerical results: the solutions to Equations \eqref{eliashberg1} and \eqref{eliashberg2}. The difficulty is that $V_{\ell'}(i\Omega)$ diverges at small $\Omega$, which leads to a divergence of $Z(i\omega)$. To deal with this numerically, we self-consistently introduce a cutoff at the scale of the putative physical gap, $\Delta_{\ell'}(0)$. This regularizes $Z(i\omega)$ and allows the coupled equations to be numerically solved.

\begin{figure}[h!]
  \centering
    \includegraphics[scale=0.6]{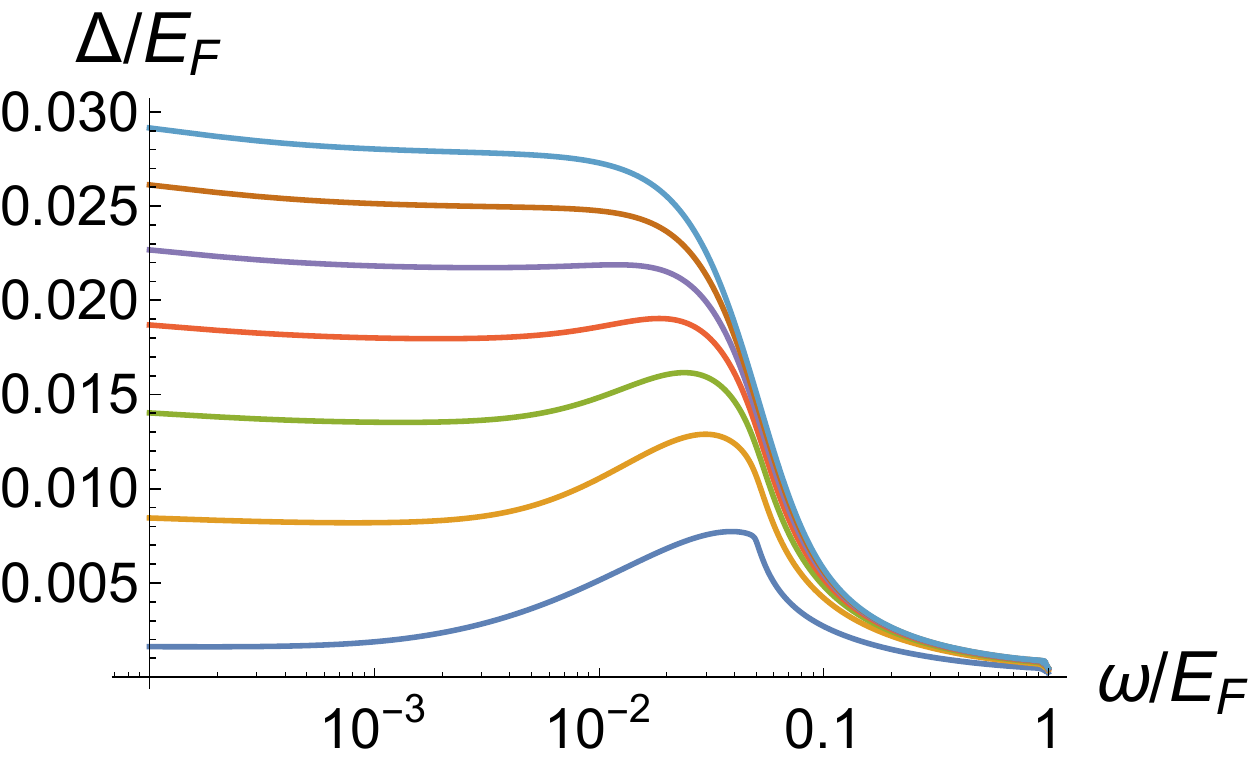}
  \caption{Gap vs Matsubara frequency for $\ell'=2$. From top to bottom: $\alpha=20,18,16,14,12,10,8$}
  \label{fig:diracCFresults}
\end{figure}
\begin{figure}[h!]
  \centering
    \includegraphics[scale=0.8]{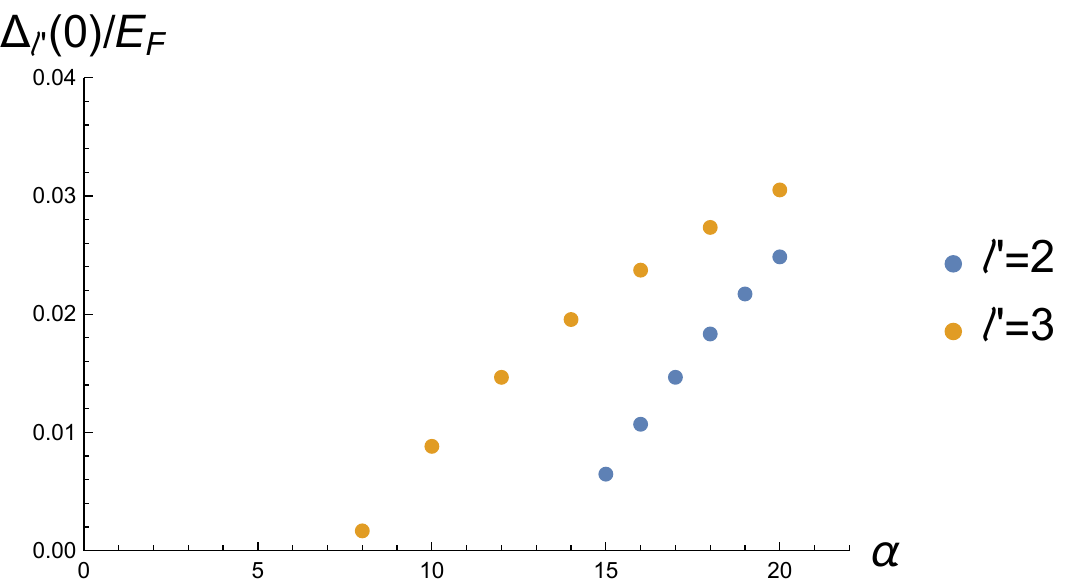}
  \caption{Physical gap vs coupling contstant for $\ell'=2,3$ pairing channels.}
  \label{fig:gaps}
\end{figure}
It is clear from Figure \ref{fig:diracCFresults} that a finite value of $\Delta_{\ell'}(0)$ is attained for large enough coupling, for $\ell'\geq2$. For $\ell'=1$, which corresponds to pairing of Dirac CFs in the $\ell=0$ mode, the potential is repulsive at all Matsubara frequencies, and thus pairing in this channel is not possible with our pairing mechanism. The results are very similar to those in Ref.~\onlinecite{Wang:2016bh}. The $T=0$ superconducting transitions appear as quantum critical points.

\subsection{Specific Heat}
We would now like to compute the low-temperature specific heat for the Dirac CFs, including the effects of current-current interactions mediated by the exchange of transverse bosons. To do this, we follow the procedure of Ref.\onlinecite{Holstein:1973zz} and use the formula of Luttinger \cite{Luttinger:1960zz} connecting the thermodynamic potential at low temperature to the diagramatically accesseble fermion propagator:
\begin{equation}{\label{luttinger}}
\begin{split}
  \Omega(T)=-V\Tr_s\int\frac{d^2\tbf{p}}{(2\pi)^2}\frac{1}{2\pi i}\int_{-\infty}^\infty & dx\{\ln[G^{-1}(p,x-i\epsilon)] \\
  &-\text{c.c.}\}\frac{1}{e^{\beta(x-\mu)}+1},
\end{split}
\end{equation}
where the $\Tr_s$ traces over the pseudospin degrees of freedom, and $V$ is the system volume. The bare fermion propagator is given by
\begin{equation}
  [G^{-1}]^{(0)}(p,i\omega)=i\vec\gamma\cdot\tbf{p}+(i\omega+\mu)\gamma_0,
\end{equation}
and the full propagator, including the fermion self-energy, can be written as
\begin{widetext}
\begin{equation}{\label{fullG}}
  G^{-1}(p,i\omega)=i\vec\gamma\cdot\hat{p}(|\tbf{p}|+\Sigma'(p,i\omega)) 
  +(i\omega+\mu+\Sigma''(p,i\omega))\gamma_0.
\end{equation}
It will be convenient to integrate by parts in $p$, and, differentiating with respect to $T$ to obtain the specific heat, we find
\begin{equation}
  c(T)=\frac{1}{8\pi^3 i}\Tr_s\int_0^{2\pi}d\theta\int_0^\infty p^2dp\int_{-\infty}^\infty dy\{G(p,\mu+yT-i\epsilon)\frac{\partial G^{-1}(p,\mu+yT-i\epsilon)}{\partial p}-\text{c.c.}\}\frac{y^2e^{y}}{(e^{y}+1)^2}.
\end{equation}
\end{widetext}
It will happen that in the region we are interested in, $\Sigma'(p,i\omega)=-\Sigma''(p,i\omega)\equiv\Sigma(p,i\omega)$. Then we can perform the angular integration and the pseudospin trace, and drop a term corresponding to degrees of freedom in the negative-energy band, to find
\begin{equation}
\begin{split}
  c(T)=&\frac{1}{8\pi^2i}\int_{-\infty}^\infty dy\frac{ye^{y}}{(e^{y}+1)^2}\int_0^\infty dp\,p^2 \\
  &\times\left\{\frac{1+2\frac{\partial\Sigma(p,\mu+yT-i\epsilon)}{\partial p}}{p+2\Sigma(p,\mu+yT-i\epsilon)-yT+i\epsilon}-\text{c.c.}\right\}.
\end{split}
\end{equation}
The integral over $p$ can be written as a contour integral:
\begin{equation}
  \int_0^\infty p^2dp\,(\cdots)=\int\oln{p}^2(z)\left(\frac{dz}{z}-\frac{d\oln{z}}{\oln{z}}\right),
\end{equation}
where $z\equiv p+2\Sigma(p,\mu+yT-i\epsilon)-yT$, and $\oln{p}(z)$ is the solution of $[p+2\Sigma(p,\mu+yT-i\epsilon)-yT]_{p=\oln{p}(z)}=z$. The contour of integration is that for which $\oln{p}(z)$ is real. Along this contour, $\im z=2\im\Sigma(p,\mu+yT-i\epsilon)$. In general, this is of order $\mathcal{O}(T^2)$; however, we show that near $z=0$, the behavior is instead of $\mathcal{O}(T)$. Thus as $T\rightarrow0$, we can approximate $\oln{p}^2(z)=\oln{p}^2(\oln{z})-4i\oln{p}(\oln{z})\im(\oln{z})\frac{d\oln{p}(\oln{z})}{d\oln{z}}$, so that
\begin{equation}
\begin{split}
  \int_0^\infty p^2dp\,(\cdots)=&\int\left(\oln{p}^2(z)\frac{dz}{z}-\oln{p}^2(\oln{z})\frac{d\oln{z}}{\oln{z}}\right) \\
  &+4i\re\int\frac{p\,\im z(p)}{z(p)}dp.
\end{split}
\end{equation}
It is shown in Appendix B that the contribution of the second integral above is subleading, and we subsequently drop it. The first integral follows a contour from left to right just above the real $z$-axis, and returns from right to left just below. Since the distance from the contour to the axis behaves as $\mathcal{O}(T)$ near $z=0$ and as $\mathcal{O}(T^2)$ away from $z=0$, we can ``pinch off'' the contour into a clockwise contour encircling $z=0$:
\begin{equation}
  \int_0^\infty p^2dp\,(\cdots)=\oint\oln{p}^2(z)\frac{dz}{z}=2\pi i\,\oln{p}^2(0).
\end{equation}
Thus we have
\begin{equation}
  c(T)=\frac{1}{4\pi}\int_{-\infty}^\infty dy\frac{ye^{y}}{(e^{y}+1)^2}\oln{p}^2(0).
\end{equation}
To evaluate this further, we need the solution of $p+2\Sigma(p,\mu+yT-i\epsilon)-yT=0$. It is shown in Appendix {B} that this quantity has the leading behavior
\begin{equation}
  \lim_{\xi\rightarrow\mu}p(\xi)=k_F-\frac{1}{\pi^2(v_F^*)^2\alpha'}(\xi-\mu)\ln|\xi-\mu|.
\end{equation}
Thus, to leading order (where $v_{F}^{*}$ is the renormalized Fermi velocity)
\begin{equation}
\begin{split}
  c(T)=\frac{1}{4\pi}&\int_{-\infty}^\infty dy\frac{ye^{y}}{(e^{y}+1)^2} \\
  &\times\left(k_F^2-\frac{2k_F}{\pi^2(v_F^*)^2\alpha'}yT\ln(yT)\right),
\end{split}
\end{equation}
or
\begin{equation}
  c(T)=-\frac{k_F}{6\pi(v_F^*)^2\alpha'}T\ln T.
\end{equation}

\section{Conclusions}
We have shown that the exchange of transverse bosons can provide a pairing mechanism for Dirac CFs, allowing for the possibility of superconductivity in the half-filled Landau level, for angular momentum channels $|\ell|\geq1$.  Previous work\cite{Wang:2016bh} could be criticized on three grounds: (a) the  wave function renormalization (the Eliashberg-$Z$ factor) was set to unity on the grounds that as long as there was a gap, the qualitative phase diagram for quantum criticality could not be changed except perhaps close to the quantum critical point. (b) Therefore only one of the two Eliashberg equations was solved. It is now clear that qualitative results remain unchanged with insignificant numerical differences. (c) The earlier work did not include the vertex correction. This could  cast doubt on our results for the superconducting transitions at $T=0$. Now we have shown that to a good approximation the vertex correction is identically zero, a far better situation than even in the electron-phonon problem. After  all these corrections taken into account, we have shown that our previous results remain semiquantitatively correct, and there is no sign of pairing in the angular momentum channel $\ell=0$.
For the specific heat, our result of $c(T)\sim \frac{1}{e^2}T\ln T$ agrees strikingly with the result of Ref.~\onlinecite{Halperin:1993tt}; thus the specific heat cannot distinguish between Son's Dirac CF theory and HLR theory. As a by product the calculated self energy can be utilized in future work.

\section{Acknowledgement}
We would like to thank Michael Mulligan for discussion. This work was supported in part by funds from David S. Saxon Presidential Term Chair at UCLA.
\appendix
\section{Effective interaction}\label{A}
This section follows Ref.~\onlinecite{Wang:2016bh} closely. In order to investigate possible pairing of Dirac CFs mediated by the exchange of the gauge bosons, we must write down a kinetic term for the emergent gauge field. There are two possible terms: a Maxwell term, $S_{\text{max}}\sim F_{\mu\nu}F^{\mu\nu}$, with $F_{\mu\nu}\equiv\partial_\mu a_\nu-\partial_\nu a_\mu$, and a term induced by the Coulomb interaction between the \emph{physical} electrons (see Equation \eqref{electronDensity}) $S_C=\frac{e^2}{\epsilon_r}\frac{\rho'_e(\tbf{x}_1)\rho'_e(\tbf{x}_2)}{|\tbf{x}_1-\tbf{x}_2|}$, where $\epsilon_r$ is the dielectric constant of the background material. 

In momentum space, we see that $S_{\text{max}}\sim |\tbf{k}|^2$ and $S_C\sim \tbf{k}$; thus the low-energy dynamics will be dominated by the Coulomb term, and that is the term we will keep. Using the Coulomb gauge, the momentum-space Coulomb action becomes
\begin{equation}{\label{coulombAction}}
  S_{C}=\frac{1}{2}\frac{e^2}{8\pi\epsilon_r}\int\frac{d\Omega d^2\tbf{k}}{(2\pi)^3}a_T(k)|\tbf{k}|a_T(-k),
\end{equation}
where we have Wick rotated so that $\Omega$ are zero-temperature Matsubara frequencies, $k\equiv(i\Omega,\tbf{k})$, and $a_T(\tbf{k})\equiv\epsilon_{ij}\hat k_ia_j(\tbf{k})$ is the transverse component of the gauge field. We see that the bare transverse gauge field propagator takes the form
\begin{equation}{\label{bareGaugeProp}}
  D_T^{(0)}(k)=\frac{8\pi\epsilon_r}{e^2}\frac{1}{|\tbf{k}|}.
\end{equation}
We can now integrate out the transverse gauge field to obtain a current-current interaction:
\begin{equation}
  S_{\text{int}}=\frac{1}{2}\int\frac{d\Omega d^2\tbf{k}}{(2\pi)^3}J_T(k)D_T^{(0)}(k)J_T(-k),
\end{equation}
with the transverse CF current operator given by $J_T(k)=\epsilon_{ij}\hat{k}_i\,i\int\frac{d\omega d^2\tbf{q}}{(2\pi)^3}\oln\psi(q+k)\gamma_i\psi(q)$. Since ($\ref{effectiveAction}$) is a low-energy effective action, it must be valid only near the Fermi surface, and so we must project this interaction to the Fermi surface. To achieve this, we make the replacement \cite{Kachru:2015ew}
\begin{equation}
  \psi(k)\rightarrow P^{(+)}_{\tbf{k}}\psi(k)=\frac{1}{\sqrt{2}}\begin{pmatrix}
    ie^{-\theta_{\tbf{k}}} \\
    1
  \end{pmatrix}\chi(k),
\end{equation}
where $P^{(+)}_{\tbf{k}}\equiv\frac{1}{2}(1+i\gamma_0\vec\gamma\cdot\hat{k})$ is the projection operator onto the positive energy branch of the Dirac CF. This gives us an interaction between scalar fields $\chi(k)$,
\begin{widetext}
\begin{equation}{\label{scalarInt}}
  S_{\text{int}}=\frac{1}{2}\int\prod_{i=1}^4\frac{d\omega_id^2\tbf{k}_i}{(2\pi)^3}(2\pi)^3\delta^{(3)}(k_3+k_4-k_2-k_1)\frac{8\pi\epsilon_r}{e^2}\frac{e^{-\frac{i}{2}[\theta_{\tbf{k}_1}+\theta_{\tbf{k}_2}-\theta_{\tbf{k}_3}-\theta_{\tbf{k}_4}]}}{|\tbf{k}_3-\tbf{k}_1|}\chi^\dagger(k_4)\chi^\dagger(k_2)\chi(k_3)\chi(k_1).
\end{equation}
\end{widetext}
We now consider this interaction in the BCS channel, $k_1=-k_2\equiv k\equiv(\omega,\tbf{k})$ and $k_3=-k_4\equiv k'\equiv(\omega',\tbf{k}')$, and define the momentum and frequency transfers $\Omega\equiv\omega'-\omega$ and $\tbf{q}\equiv\tbf{k}'-\tbf{k}$. Then, making the Fermi surface approximation $|\tbf{k}|=|\tbf{k}'|= k_F$, we can read off an effective BCS-channel interaction
\begin{equation}
  V_{\text{BCS}}(\tbf{k},\tbf{k}')=\frac{4\pi\epsilon_r}{k_Fe^2}\frac{e^{-i[\theta_{\tbf{k}}-\theta_{\tbf{k}'}]}}{|\sin\frac{\theta_{\tbf{k}}-\theta_{\tbf{k}'}}{2}|}.
\end{equation}
In order to generate an attractive interaction, we introduce an RPA-corrected potential, with a correction from screening due to the finite density of Dirac CFs; and, integrating over the Fermi surface in the $\ell'$ angular momentum channel, we generate an effective interaction
\begin{equation}
  V_{\ell'}(i\Omega)\equiv\alpha\int\frac{d\theta}{2\pi}\frac{e^{i(\ell'-1)\theta}}{|\sin\frac{\theta}{2}|}\frac{2}{1+\alpha\frac{|\Omega|}{|\sin\frac{\theta}{2}|}}.
\end{equation}
Here we have an effective coupling constant $\alpha\equiv\frac{\epsilon_r}{e^2}$. This $\ell'$ is the angular momentum channel for the scalar field $\chi(k)$; its relationship to $\ell$, the angular momentum channel of the Dirac CF, depends on the nature of the order parameter $\hat\Delta(k)\equiv[\Delta_s(k)+\tbf{d}(k)\cdot\bm{\sigma}]i\sigma_2$. For the pseudospin singlet order parameter, $\hat\Delta(k)=\langle\psi^T(-k)P^{(+)}_{-\tbf{k}}i\sigma_2P_{\tbf{k}}^{(+)}\psi(k)\rangle$, $\ell=\ell'-1$, and in order to satisfy antisymmetry of $\hat\Delta(k)$, $\ell$ must be even. For the pseudospin triplet, $\hat\Delta(k)=\langle\psi^T(-k)P^{(+)}_{-\tbf{k}}(\tbf{d}\cdot\bm{\sigma}) (i\sigma_2)P_{\tbf{k}}^{(+)}\psi(k)\rangle$, $\ell$ must be odd, and either $\ell=\ell'$ or $\ell=\ell'-2$, depending on which triplet state the pair is in.

\section{Fermion self-energy}\label{B}
\begin{figure}
  \includegraphics[scale=0.3]{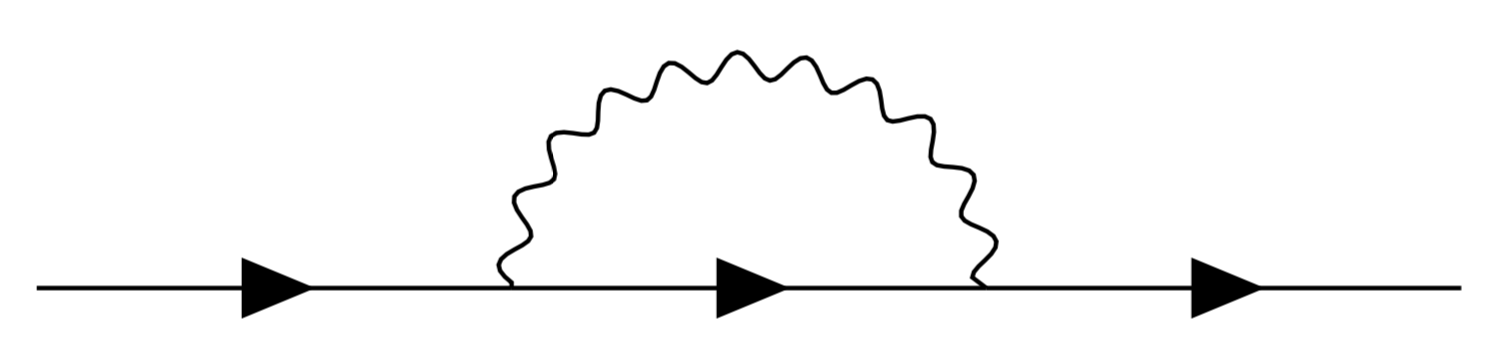}
  \caption{\label{oneloop}One-loop correction to fermion propagator}
\end{figure}
We would like to compute the one-loop Dirac CF self-energy, diagrammatically shown in Figure \ref{oneloop}. Each vertex gives a factor of $i\gamma_i$, and we will use a corrected version of the boson propagator that takes into account screening from the finite density of fermions:
\begin{equation}
  D(k,i\omega)P^T_{ij}(k)=\frac{8\pi}{\alpha'}\frac{1}{k+\frac{2k_F}{\alpha'}\frac{|\omega|}{k}}P^T_{ij}(k) ,
\end{equation}
where $\alpha'\equiv\frac{e^2}{\epsilon_r}$ is the coupling of the original electrons in the problem, $P^T_{ij}(k)\equiv\delta_{ij}-\hat{k}_i\hat{k}_j$ is the transverse projector, and $k\equiv|\tbf{k}|$. Importantly, the bosons are unscreened at small $\omega/k$; this leads to the anomalous behavior in the specific heat. We represent the fermion propagator by splitting it up into positive- and negative-energy parts as (see, e.g., Ref.~\onlinecite{Miransky:2001jba})
\begin{equation}
  G(\tbf{k},i\omega_n)=\gamma_0\sum_{s=\pm}G^{(s)}(\tbf{k},i\omega_n)P^{(s)}_{\tbf{k}},
\end{equation}
where $P^{(s)}_{\tbf{k}}\equiv\frac{1}{2}(1+si\gamma_0\vec\gamma\cdot\hat{k})$ is the projector onto the positive or negative energy bands, and
\begin{equation}
  G^{(s)}(\tbf{k},i\omega_n)=\frac{1}{i\omega_n-s|\tbf{k}|+\mu}
\end{equation}

Then the relevant diagram gives the contribution
\begin{equation}
\begin{split}
  \Sigma(\tbf{k},i\nu_n)&=\frac{1}{\beta^2}\sum_{r,m}\sum_{s=\pm}\int_p(i\gamma_i)\gamma_0P^{(s)}_{\tbf{k}}G^{(s)}(\tbf{k},i\xi_r) \\
  &\times D(p-k,iz_m)P^T_{ij}(p)(\beta\delta_{\xi_r+z_m,\nu_n})(i\gamma_j).
\end{split}
\end{equation}
Performing the sum over Matsubara frequencies, working out the matrix structure, and analytically continuing $i\nu_n\rightarrow\nu+i\epsilon$, we find that 
\begin{equation}
\begin{split}
  \im\Sigma'_R(\tbf{k},\nu)=-\frac{1}{2\pi}\sum_{s=\pm}s&\int_p\int_{\mu}^\nu d\xi'\,\im G_R^{(s)}(p,\xi') \\
  &\times\im D_R(k-p,\nu-\xi'),
\end{split}
\end{equation}
where $\Sigma'$ is defined in Equation \eqref{fullG}. Now, we are interested in evaluating the self-energy on the Fermi surface, i.e. $|\tbf{k}|\approx k_F$, $\nu\approx\mu$. In this limit, the region of frequency integration above is squeezed around $\xi'\approx\mu$, and we can simplify the fermion propagator:
\begin{equation}
  \lim_{\xi'\rightarrow\mu}\im G_R^{(s)}(p,\xi')=-\frac{\pi}{v_F^*}\delta(p-p(\xi'))\delta_{s,+},
\end{equation}
where $v_F^*\equiv|1+\frac{\partial}{\partial p}\Sigma_R(p,\xi')|_{k_F}$, and $p(\xi')$ is defined as the solution to 
\begin{equation}{\label{pxiDef}}
  \xi'+\mu-p-2\re\Sigma'_R(\tbf{p},\xi')=0,
\end{equation}
i.e. $p(\xi')$ is the momentum at the pole of the positive energy branch of the fermion propagator. Note that we have taken $\mu>0$, and so the $s=-$ portion of the propagator does not contribute. With this substitution, it becomes clear that $\Sigma''_R=-\Sigma'_R$, and we now define $\Sigma_R\equiv\Sigma'_R$. Using this simplification, we can perform the angular integration of $\tbf{p}$, to obtain
\begin{equation}
\begin{split}
  \im\Sigma_R(\tbf{k},\nu)=\frac{1}{8\pi^2v_F^*}\frac{1}{k}&\int_{\mu}^\nu d\xi'p(\xi')\int_{|p(\xi')-k|}^{p(\xi')+k}dp \\
  &\frac{\im D_R(p,\nu-\xi')}{\sqrt{1-\left[\frac{p^2+k^2-p^2(\xi')}{2pk}\right]^2}}.
\end{split}
\end{equation}
Then, keeping only leading-order terms, these integrals can be performed, to obtain eventually
\begin{equation}{\label{imSigma}}
\begin{split}
  \im\Sigma_R(\tbf{k},\nu)=-\frac{1}{2\pi v_F^*\alpha'}&(\nu-\mu) \\
  &\times\tan^{-1}\left(\frac{\frac{2k_F}{\alpha'}(\nu-\mu)}{(k-k_F)^2}\right).
\end{split}
\end{equation}
Here the limits $k\approx k_F$ and $\nu\approx\mu$ are understood. This expression contains the anomalous behavior on the Fermi surface.

We next show that this behavior of $\im\Sigma_R$ leads to a logarithmic divergence of $\frac{\partial}{\partial\xi}\re\Sigma_R(p,\xi)$ on the Fermi surface. We achieve this by means of the Kramers-Kronig relations, which give us, after an integration by parts,
\begin{equation}
  \frac{\partial}{\partial\xi}\re\Sigma_R(\tbf{k},\xi)=\frac{\mathcal{P}}{\pi}\int_{-\infty}^\infty d\xi'\frac{\frac{\partial}{\partial\xi'}\im\Sigma_R(\tbf{k},\xi')}{\xi'-\xi}.
\end{equation}
Substituting in Equation \eqref{imSigma}, and taking the principal part of the integral, we find that (up to finite terms)
\begin{equation}{\label{partialReSigma}}
  \frac{\partial}{\partial\xi}\re\Sigma_R(\tbf{k},\xi)=\frac{1}{2\pi v_F^*\alpha'}\ln(\xi-\mu).
\end{equation}
Now, differentiating Equation \eqref{pxiDef}, and substituting in Equation \eqref{partialReSigma}, we see that
\begin{equation}
  \frac{dp(\xi)}{d\xi}=\frac{1}{v_F^*}\left(1-\frac{1}{\pi v_F^*\alpha'}\ln(\xi-\mu)\right).
\end{equation}
Finally, we can integrate this and drop subleading terms to obtain
\begin{equation}{\label{pxi}}
  p(\xi)=k_F-\frac{1}{\pi (v_F^*)^2\alpha'}(\xi-\mu)\ln(\xi-\mu).
\end{equation}

\section{Vertex correction}\label{C}
In order for the Eliashberg equations to be trustworthy, the vertex corrections must be negligible. Here we can make use of the Ward identity,~\cite{Chakravarty1995} which gives us the vertex correction in terms of the self-energy:
\begin{equation}{\label{ward}}
  \hat{\tbf{k}}\cdot\nabla_{\tbf{k}}\Sigma(k,\mu)|_{k_F}=\Gamma^{(2)}(k_F,k_F,\mu).
\end{equation}
Using Equation \eqref{imSigma}, we can compute the derivative of the self-energy:
\begin{equation}
  \im\hat{\tbf{k}}\cdot\nabla_{\tbf{k}}(k,\nu)\propto\frac{(\nu-\mu)^2(k-k_F)}{\left(\frac{2k_F(\nu-\mu)}{\alpha'}\right)^2+(k-k_F)^4}.
\end{equation}
This vanishes on the Fermi surface, so that $\im\Gamma^{(2)}(k_F,k_F,\nu)=0$. Similarly, using the Kramers-Kronig relations to obtain the real part of the self-energy, we also find that $\re\Gamma^{(2)}(k_F,k_F,\nu)=0$, so that
\begin{equation}
  \Gamma^{(2)}(k_F,k_F,\mu)=0.
\end{equation}
This justifies the use of the Eliashberg equations in this problem, and also justifies the use of Luttinger's expansion of the thermodynamic potential, Equation \eqref{luttinger}.


\begin{thebibliography}{14}%
\makeatletter
\providecommand \@ifxundefined [1]{%
 \@ifx{#1\undefined}
}%
\providecommand \@ifnum [1]{%
 \ifnum #1\expandafter \@firstoftwo
 \else \expandafter \@secondoftwo
 \fi
}%
\providecommand \@ifx [1]{%
 \ifx #1\expandafter \@firstoftwo
 \else \expandafter \@secondoftwo
 \fi
}%
\providecommand \natexlab [1]{#1}%
\providecommand \enquote  [1]{``#1''}%
\providecommand \bibnamefont  [1]{#1}%
\providecommand \bibfnamefont [1]{#1}%
\providecommand \citenamefont [1]{#1}%
\providecommand \href@noop [0]{\@secondoftwo}%
\providecommand \href [0]{\begingroup \@sanitize@url \@href}%
\providecommand \@href[1]{\@@startlink{#1}\@@href}%
\providecommand \@@href[1]{\endgroup#1\@@endlink}%
\providecommand \@sanitize@url [0]{\catcode `\\12\catcode `\$12\catcode
  `\&12\catcode `\#12\catcode `\^12\catcode `\_12\catcode `\%12\relax}%
\providecommand \@@startlink[1]{}%
\providecommand \@@endlink[0]{}%
\providecommand \url  [0]{\begingroup\@sanitize@url \@url }%
\providecommand \@url [1]{\endgroup\@href {#1}{\urlprefix }}%
\providecommand \urlprefix  [0]{URL }%
\providecommand \Eprint [0]{\href }%
\providecommand \doibase [0]{http://dx.doi.org/}%
\providecommand \selectlanguage [0]{\@gobble}%
\providecommand \bibinfo  [0]{\@secondoftwo}%
\providecommand \bibfield  [0]{\@secondoftwo}%
\providecommand \translation [1]{[#1]}%
\providecommand \BibitemOpen [0]{}%
\providecommand \bibitemStop [0]{}%
\providecommand \bibitemNoStop [0]{.\EOS\space}%
\providecommand \EOS [0]{\spacefactor3000\relax}%
\providecommand \BibitemShut  [1]{\csname bibitem#1\endcsname}%
\let\auto@bib@innerbib\@empty
\bibitem [{\citenamefont {Jiang}\ \emph {et~al.}(1989)\citenamefont {Jiang},
  \citenamefont {Stormer}, \citenamefont {Isui}, \citenamefont {Pfeiffer},\
  and\ \citenamefont {West}}]{Jiang:1989de}%
  \BibitemOpen
  \bibfield  {author} {\bibinfo {author} {\bibfnamefont {H.~W.}\ \bibnamefont
  {Jiang}}, \bibinfo {author} {\bibfnamefont {H.~L.}\ \bibnamefont {Stormer}},
  \bibinfo {author} {\bibfnamefont {D.~C.}\ \bibnamefont {Isui}}, \bibinfo
  {author} {\bibfnamefont {L.~N.}\ \bibnamefont {Pfeiffer}}, \ and\ \bibinfo
  {author} {\bibfnamefont {K.~W.}\ \bibnamefont {West}},\ }\href@noop {}
  {\bibfield  {journal} {\bibinfo  {journal} {Physical Review B}\ }\textbf
  {\bibinfo {volume} {40}},\ \bibinfo {pages} {12013} (\bibinfo {year}
  {1989})}\BibitemShut {NoStop}%
\bibitem [{\citenamefont {Willett}\ \emph {et~al.}(1990)\citenamefont
  {Willett}, \citenamefont {Paalanen}, \citenamefont {Ruel}, \citenamefont
  {West}, \citenamefont {Pfeiffer},\ and\ \citenamefont
  {Bishop}}]{Willett:1990fn}%
  \BibitemOpen
  \bibfield  {author} {\bibinfo {author} {\bibfnamefont {R.~L.}\ \bibnamefont
  {Willett}}, \bibinfo {author} {\bibfnamefont {M.~A.}\ \bibnamefont
  {Paalanen}}, \bibinfo {author} {\bibfnamefont {R.~R.}\ \bibnamefont {Ruel}},
  \bibinfo {author} {\bibfnamefont {K.~W.}\ \bibnamefont {West}}, \bibinfo
  {author} {\bibfnamefont {L.~N.}\ \bibnamefont {Pfeiffer}}, \ and\ \bibinfo
  {author} {\bibfnamefont {D.~J.}\ \bibnamefont {Bishop}},\ }\href@noop {}
  {\bibfield  {journal} {\bibinfo  {journal} {Physical Review Letters}\
  }\textbf {\bibinfo {volume} {65}},\ \bibinfo {pages} {112} (\bibinfo {year}
  {1990})}\BibitemShut {NoStop}%
\bibitem [{\citenamefont {Du}\ \emph {et~al.}(1994)\citenamefont {Du},
  \citenamefont {Stormer}, \citenamefont {Tsui}, \citenamefont {Yeh},
  \citenamefont {Pfeiffer},\ and\ \citenamefont {West}}]{Du:1994wtw}%
  \BibitemOpen
  \bibfield  {author} {\bibinfo {author} {\bibfnamefont {R.~R.}\ \bibnamefont
  {Du}}, \bibinfo {author} {\bibfnamefont {H.~L.}\ \bibnamefont {Stormer}},
  \bibinfo {author} {\bibfnamefont {D.~C.}\ \bibnamefont {Tsui}}, \bibinfo
  {author} {\bibfnamefont {A.~S.}\ \bibnamefont {Yeh}}, \bibinfo {author}
  {\bibfnamefont {L.~N.}\ \bibnamefont {Pfeiffer}}, \ and\ \bibinfo {author}
  {\bibfnamefont {K.~W.}\ \bibnamefont {West}},\ }\href@noop {} {\bibfield
  {journal} {\bibinfo  {journal} {Physical Review Letters}\ }\textbf {\bibinfo
  {volume} {73}},\ \bibinfo {pages} {3274} (\bibinfo {year}
  {1994})}\BibitemShut {NoStop}%
\bibitem [{\citenamefont {Halperin}\ \emph {et~al.}(1993)\citenamefont
  {Halperin}, \citenamefont {Lee},\ and\ \citenamefont
  {Read}}]{Halperin:1993tt}%
  \BibitemOpen
  \bibfield  {author} {\bibinfo {author} {\bibfnamefont {B.~I.}\ \bibnamefont
  {Halperin}}, \bibinfo {author} {\bibfnamefont {P.~A.}\ \bibnamefont {Lee}}, \
  and\ \bibinfo {author} {\bibfnamefont {N.}~\bibnamefont {Read}},\ }\href@noop
  {} {\bibfield  {journal} {\bibinfo  {journal} {Physical Review B}\ }
  (\bibinfo {year} {1993})}\BibitemShut {NoStop}%
\bibitem [{\citenamefont {Son}(2015)}]{Son:2015gz}%
  \BibitemOpen
  \bibfield  {author} {\bibinfo {author} {\bibfnamefont {D.~T.}\ \bibnamefont
  {Son}},\ }\href@noop {} {\bibfield  {journal} {\bibinfo  {journal} {Physical
  Review X}\ }\textbf {\bibinfo {volume} {5}},\ \bibinfo {pages} {031027}
  (\bibinfo {year} {2015})}\BibitemShut {NoStop}%
\bibitem [{\citenamefont {Levin}\ and\ \citenamefont
  {Son}(2017)}]{Levin:2017fk}%
  \BibitemOpen
  \bibfield  {author} {\bibinfo {author} {\bibfnamefont {M.}~\bibnamefont
  {Levin}}\ and\ \bibinfo {author} {\bibfnamefont {D.~T.}\ \bibnamefont
  {Son}},\ }\href@noop {} {\bibfield  {journal} {\bibinfo  {journal} {Physical
  Review B}\ }\textbf {\bibinfo {volume} {95}},\ \bibinfo {pages} {125120}
  (\bibinfo {year} {2017})}\BibitemShut {NoStop}%
\bibitem [{\citenamefont {Wang}\ \emph {et~al.}(2017)\citenamefont {Wang},
  \citenamefont {Cooper}, \citenamefont {Halperin},\ and\ \citenamefont
  {Stern}}]{Wang:2017kx}%
  \BibitemOpen
  \bibfield  {author} {\bibinfo {author} {\bibfnamefont {C.}~\bibnamefont
  {Wang}}, \bibinfo {author} {\bibfnamefont {N.~R.}\ \bibnamefont {Cooper}},
  \bibinfo {author} {\bibfnamefont {B.~I.}\ \bibnamefont {Halperin}}, \ and\
  \bibinfo {author} {\bibfnamefont {A.}~\bibnamefont {Stern}},\ }\href@noop {}
  {\bibfield  {journal} {\bibinfo  {journal} {Physical Review X}\ }\textbf
  {\bibinfo {volume} {7}},\ \bibinfo {pages} {590} (\bibinfo {year}
  {2017})}\BibitemShut {NoStop}%
\bibitem [{\citenamefont {Wang}\ and\ \citenamefont
  {Chakravarty}(2016)}]{Wang:2016bh}%
  \BibitemOpen
  \bibfield  {author} {\bibinfo {author} {\bibfnamefont {Z.}~\bibnamefont
  {Wang}}\ and\ \bibinfo {author} {\bibfnamefont {S.}~\bibnamefont
  {Chakravarty}},\ }\href@noop {} {\bibfield  {journal} {\bibinfo  {journal}
  {Physical Review B}\ }\textbf {\bibinfo {volume} {94}},\ \bibinfo {pages}
  {165138} (\bibinfo {year} {2016})}\BibitemShut {NoStop}%
\bibitem [{\citenamefont {Luttinger}(1960)}]{Luttinger:1960zz}%
  \BibitemOpen
  \bibfield  {author} {\bibinfo {author} {\bibfnamefont {J.~M.}\ \bibnamefont
  {Luttinger}},\ }\href@noop {} {\bibfield  {journal} {\bibinfo  {journal}
  {Phys. Rev.}\ }\textbf {\bibinfo {volume} {119}},\ \bibinfo {pages} {1153}
  (\bibinfo {year} {1960})}\BibitemShut {NoStop}%
\bibitem [{\citenamefont {Curnoe}\ and\ \citenamefont
  {Stamp}(1998)}]{Curnoe:1998fo}%
  \BibitemOpen
  \bibfield  {author} {\bibinfo {author} {\bibfnamefont {S.}~\bibnamefont
  {Curnoe}}\ and\ \bibinfo {author} {\bibfnamefont {P.~C.~E.}\ \bibnamefont
  {Stamp}},\ }\href@noop {} {\bibfield  {journal} {\bibinfo  {journal}
  {Physical Review Letters}\ }\textbf {\bibinfo {volume} {80}},\ \bibinfo
  {pages} {3312} (\bibinfo {year} {1998})}\BibitemShut {NoStop}%
\bibitem [{\citenamefont {Holstein}\ \emph {et~al.}(1973)\citenamefont
  {Holstein}, \citenamefont {Norton},\ and\ \citenamefont
  {Pincus}}]{Holstein:1973zz}%
  \BibitemOpen
  \bibfield  {author} {\bibinfo {author} {\bibfnamefont {T.}~\bibnamefont
  {Holstein}}, \bibinfo {author} {\bibfnamefont {R.~E.}\ \bibnamefont
  {Norton}}, \ and\ \bibinfo {author} {\bibfnamefont {P.}~\bibnamefont
  {Pincus}},\ }\href@noop {} {\bibfield  {journal} {\bibinfo  {journal} {Phys.
  Rev.}\ }\textbf {\bibinfo {volume} {B8}},\ \bibinfo {pages} {2649} (\bibinfo
  {year} {1973})}\BibitemShut {NoStop}%
\bibitem [{\citenamefont {Kachru}\ \emph {et~al.}(2015)\citenamefont {Kachru},
  \citenamefont {Mulligan}, \citenamefont {Torroba},\ and\ \citenamefont
  {Wang}}]{Kachru:2015ew}%
  \BibitemOpen
  \bibfield  {author} {\bibinfo {author} {\bibfnamefont {S.}~\bibnamefont
  {Kachru}}, \bibinfo {author} {\bibfnamefont {M.}~\bibnamefont {Mulligan}},
  \bibinfo {author} {\bibfnamefont {G.}~\bibnamefont {Torroba}}, \ and\
  \bibinfo {author} {\bibfnamefont {H.}~\bibnamefont {Wang}},\ }\href@noop {}
  {\bibfield  {journal} {\bibinfo  {journal} {Physical Review B}\ }\textbf
  {\bibinfo {volume} {92}},\ \bibinfo {pages} {235105} (\bibinfo {year}
  {2015})}\BibitemShut {NoStop}%
\bibitem [{\citenamefont {Miransky}\ \emph {et~al.}(2001)\citenamefont
  {Miransky}, \citenamefont {Semenoff}, \citenamefont {Shovkovy},\ and\
  \citenamefont {Wijewardhana}}]{Miransky:2001jba}%
  \BibitemOpen
  \bibfield  {author} {\bibinfo {author} {\bibfnamefont {V.~A.}\ \bibnamefont
  {Miransky}}, \bibinfo {author} {\bibfnamefont {G.~W.}\ \bibnamefont
  {Semenoff}}, \bibinfo {author} {\bibfnamefont {I.~A.}\ \bibnamefont
  {Shovkovy}}, \ and\ \bibinfo {author} {\bibfnamefont {L.~C.~R.}\ \bibnamefont
  {Wijewardhana}},\ }\href@noop {} {\bibfield  {journal} {\bibinfo  {journal}
  {Phys. Rev. D}\ }\textbf {\bibinfo {volume} {64}},\ \bibinfo {pages} {390}
  (\bibinfo {year} {2001})}\BibitemShut {NoStop}%
\bibitem [{\citenamefont {Chakravarty}\ \emph {et~al.}(1995)\citenamefont
  {Chakravarty}, \citenamefont {Norton},\ and\ \citenamefont
  {Sylju{\aa}sen}}]{Chakravarty1995}%
  \BibitemOpen
  \bibfield  {author} {\bibinfo {author} {\bibfnamefont {S.}~\bibnamefont
  {Chakravarty}}, \bibinfo {author} {\bibfnamefont {R.~E.}\ \bibnamefont
  {Norton}}, \ and\ \bibinfo {author} {\bibfnamefont {O.~F.}\ \bibnamefont
  {Sylju{\aa}sen}},\ }\href@noop {} {\bibfield  {journal} {\bibinfo  {journal}
  {Phys. Rev. Lett.}\ }\textbf {\bibinfo {volume} {74}} (\bibinfo {year}
  {1995})}\BibitemShut {NoStop}%
\end{thebibliography}
\end{document}